# Project-based physics labs using low-cost open-source hardware

F. Bouquet* and J. Bobroff

Laboratoire de Physique des Solides, CNRS, Univ. Paris-Sud, Université Paris-Saclay, 91405 Orsay Cedex, France

M. Fuchs-Gallezot and L. Maurines

Laboratoire DidaScO, Univ. Paris-Sud, Université Paris-Saclay, 91405 Orsay cedex, France

We describe a project-based physics lab, which we proposed to third-year university students. Theses labs are based on new open-source low-cost equipment (Arduino microcontrollers and compatible sensors). Students are given complete autonomy: they develop their own experimental setup and study the physics topic of their choice. The goal of these projects is to let students discover the reality of experimental physics. Technical specifications of the acquisition material and case studies are presented for practical implementation in other universities.

## I. INTRODUCTION

Traditionally, student labs are used in physics curricula to let the students discover and measure phenomena they are otherwise studying. Experimental setups can range from very low tech (a stopwatch to measure the fall of a rock) to elaborate high tech (research-lab setup), but a key parameter for a successful learning is student engagement.

Recently, the use of microcontrollers has been much simplified by the development of the famous Arduino microcontroller. This open-source low-cost microcontroller is widely used by the maker community.[1] From a technical point of view, these boards can be used as a low-cost data acquisition card. At the university level Arduino is gaining popularity: for example workshops targeting teachers and promoting these boards in advanced labs have been organized.[2] Many student labs have been rethought using this technology. Using Arduino boards allows students to build low-cost setups,[3-6] such as a computerized mirror system for optical setups,[3] or a giant stopwatch and datalogger.[5]

The low cost and flexibility of Arduino are not its only advantages: its open-source fablab nature can encourage sharing of ideas, tinkering and creativity among students. In terms of pedagogy, such an engaging environment is ideally suited to a project-based learning (PBL) framework.[7-12] Many PBL examples reported in the literature were implemented in high-school.[11,13] In comparison, fewer cases of project-based student labs have been reported at the university level.[7,8,14-18] Some rare universities have fully integrated PBL as the core of their pedagogy.[19,20] Several parameters reduce the appeal of a project-based approach in physics curricula at university level: it generally requires a large set of versatile and often expensive equipment; it requires more time than traditional teachings, and for the instructors PBL can be destabilizing.[9,13] Several strategies can be used to downsize the cost of the equipment, such as the use of cheap electronic components,[16] or building up a stock of used lab equipment over the years.[18] The apparition of the Arduino microcontroller opens new possibilities. A low-cost microcontroller with various inputs/outputs and sensors is indeed the perfect low-cost Swiss-army knife for physics projects:[21] it gives students an easy way to acquire data with a large flexibility in terms of set-up design. However, the technical specifications of the Arduino boards present strong limitations compared to more specialized data acquisition cards, in term of digitalization and sampling rate.[1]

In this article, we describe a project-based student lab using Arduino boards to acquire data, where students build their own experimental set-up from scratch. These labs are part of a broader endeavor to renew physics teaching in our university, for instance students PBL and physics outreach.[22] The aim of this article is to present a detailed description of this course so that it can inspire other teachers, especially those interested in the PBL approach but unsure of its technical feasibility. Indeed, the questions we were facing prior to this course were whether the Arduino board was the right tool for a PBL-based student lab at university level, and whether these projects could provide students with a realistic introduction to experimental physics. To answer these questions we first describe this teaching unit, its organization and goals. We include a technical description of the acquisition material for physics experiments, its sensitivity and its cost. Students' projects are then described, with an emphasis on some examples and students' results. Finally, we report on the students' and teachers' perceptions of these projects through a survey.

## II. DESCRIPTION OF THE "OPEN-PROJECT" STUDENT-LAB TEACHING UNIT

The present Arduino open projects have been introduced in an otherwise classical academic environment. The students are in their third year of the French university Paris-Sud in a fundamental physics section. Most of them have been following a physics curriculum that relies heavily on theoretical and calculation skills with a low emphasis on student labs during their first and second years. In their third and fourth years the amount of time dedicated to lab work increases. In particular, a module of student labs (dubbed "focused labs") consists in five-day labs where students have to use an elaborate experimental setup. The setups are very specialized and focus on a single experiment and physics topic, for example a setup to measure the superconducting transition, or to measure the blackbody radiation. Students use a top-equipment experimental setup, study complicated physics phenomena, and discover the difficulties of experimental work. However, no freedom is given to the students as far as the experimental setup (and the underlying physics) is concerned.

The organization of student labs was rethought in 2014 for the whole fundamental physics section. In the course of this reorganization we developed new project-based student labs (dubbed "open projects") mirroring the organization of the focused labs. The main pedagogical objective of the open projects is to be a realistic introduction to experimental physics: even though understanding the phenomenon that is studied is important, the focus is more on how to perform a scientific study and the skills it requires, from the conception of the experiment to the analysis of the results. Students are given a complete experimental autonomy and can choose which physics topic they want to study: their task is to build an experimental setup from scratch and to carry experiments in what-ever direction they think is best. In other words, they have to lead their own research project – in five days. This approach corresponds to the "discipline project" PBL as defined by Ref. 10.

This lab course is divided in two parts. Before the project itself, a first two-day period is dedicated to students' training on the acquisition material (Arduino board and sensors). The approach is learning-by-doing: after some simple exercises to master Arduino basics, challenges are given to the students mixing arbitrary constraints and type of measurements, such as "build a game using two different kinds of sensor". Additional basic materials are available, such as lego bricks, duct tape, electronic components, aluminum foil… The objective of these two days is to let students learn how to use Arduino boards and sensors to measure physical quantities, but also to let them realize how easy it is to build things. No physics is involved at that level, except basic electronics: the main goal of this first period is to engage students with an open and creative approach.

Near the end of these training days, the students spend a couple of hours on a collective brainstorming session in which they list as many potential physics projects as possible. Based on this list, the students, working in pairs, define their project themselves: they choose the physics topic and begin to work out how to investigate it. Note that no pre-made topic list is proposed by teachers, contrary to many project-based lab teaching. Indeed, we hope that the students' motivation increases if they propose themselves the topic of their research. The teachers' only role is to validate the project in term of feasibility and check with the students for special needs (for example if a specific material is required).

The project itself takes place in a second period of five consecutive days. The students have at their disposal the same material than during the training days plus some other useful material (metallic wires and plates, plastic foils, multimeters). The students' objectives are very ambitious for only five days: they have to conceive the experimental setup, build it, test it, and measure whatever physical phenomenon they chose. They must also analyze properly their data and interpret their results. As in any project, tries and errors are expected, and the students are told to await some delays or changes in their original project. Teachers regularly come and discuss the progress of the project, mentoring the students. Except when there is a security issue (such as using a 10 A / 20 V power source without supervision), the students are free to explore any direction they want; however they have to justify their choices. The teachers provide a more extensive help for some specific technical skills (how to solder a wire, how to use the fitting software, etc.). Changes in the original project are accepted, but the students are expected to produce some measurements at the end of the five days, to be able to explain what they did and why, and to discuss the physics they measured. The assessment consists of a 15-minute oral exam and a written report.

## III. ACQUISITION MATERIALS: ARDUINO BOARDS AND SENSORS FOR PHYSICS

Letting the students decide their own research topic is in line with our objective to provide an introduction to experimental physics. A consequence of this is that a large set of diverse materials should be available. The choice of Arduino boards as the backbone of the open projects is deliberate. The use of a low-cost multi-purpose microcontroller to pilot data acquisition limits the total cost (an Arduino Uno board is about $20 and the coding interface software is open source and free). Arduino is not the only low-cost microcontroller,[23] but it is recognized as very user-friendly and its user community is large. Only basic coding skill is required to operate the Arduino board as many code snippets for various sensors are available on the Internet. In terms of connectivity, it just needs an USB port, so that students can use their own laptop and can even bring their project outside of the lab rooms if needed.

| Physical parameter | sensor | range | resolution |
|---|---|---|---|
| Voltage | Arduino board analog input | 0 V – 5 V | 5mV |
| | voltage amplifier for thermocouple MAX31855 | Typical -10mV to 40mV | 10 µV |
| | Arduino board analog output | 0 V – 5 V , limited at 40 mA | 20 mV, pulse-width modulation, |
| Magnetic field | Tinkerkit T000070 hall sensor | -2000 G to 2000 G | 4 G |
| | 3-axis HMC5883L magnetometer | -8 G to 8 G (variable) | ~ 1 mG to 5 mG |
| Temperature | Tinkerkit T000200 thermistor | not calibrated | |
| | K-thermocouple + amplifier MAX31855 | -250°C to 1300°C | 0.25 °C |
| | Arduino + Pt100 | -250°C to 100°C | typical 0.5 °C |
| Light | Tinkerkit LDR sensor | not calibrated | |
| | Sparkfun temt6000 | 0 to 1000 lux | 1 lux |
| | Adafruit TSL2561 | Dynamic range from 0.1 lux to 40000 lux | Depends on the range |
| Sound | SparkFun Electret Microphone BOB-09964 | not calibrated | |
| Acceleration | TinkerKit 2/3 Axis Accelerometer | not calibrated | |
| | ADXL335 Accelerometer | -3 g to +3 g | 0.01 g |
| Force | FSR01 force sensing resistor | 0.2 N to 20 N | Non linear |
| | Strain Gauge based CZL 616 C + amplifier MAX31855 | 0-780 g | 1 g |

Table 1: list of sensors and their specifications (most cost less than $10). This list gives an idea of the physics phenomena that can be studied. This list is given as an example: numerous other sensors exist.

Table 1 shows the specifications of the various (low-cost) sensors our students had at their disposal for their projects and the corresponding physics measurements that can be performed. The diversity of the projects is obviously linked to the variety of sensors available. These sensors don't have the sensitivity of lab-quality equipment, but they offer a wide range of physics phenomena that can be measured, and studied.

All in all, not counting the computers used for data analysis, the total cost of the material used in each student's project was less than $100. So with a total budget under $2000 and some computers, it is possible to successfully organize a physics project-based students' lab for about 20 students.

At first, the Arduino and sensors presented here were chosen mostly because of their low cost, allowing us to test this teaching with no financial risk. However, we realized that using low-cost hardware had also an influence on our pedagogy: instructors can encourage students to experiment in whatever they think is interesting, even at the risk of damag-ing the equipment. One or two Arduino boards were fried during the projects, which is a very small price for complete student autonomy. We argue that it played a role in the success of these projects, for the students as well as for the teachers.

## IV. STUDENTS' PROJECTS

Prior to the project week, the question of whether Arduino boards and low-cost sensors would allow students to perform studies of interest for a third-year university physics curriculum was open. After completion of the projects, all eleven pairs of students succeeded in producing a working experimental setup and physical measurements, even though some projects have been reoriented along their course and their ambition downsized. Table 2 lists the projects that have been carried out and it shows the diversity of the topics that were studied. Generally speaking, in five days the open-project students had time to build a setup, test it, and run a few series of measurements, even though more time would be needed to perform a complete study.

We present three typical student projects that explore different fields of physics.

### A. ELECTRICAL PROPERTIES OF MATTER

With its analog inputs and outputs, the Arduino board can be used directly to study the electrical conductivity. A simple voltage-divider circuit with a reference resistor in series allows the measurement of a sample resistor through the analog input of the board. Changing the reference resistor of the voltage divider allows scanning a large range of sample resistance, from 150 Ω to 60 kΩ in this case. The temperature can be determined with the same electrical circuit, measuring the resistance of a standard Pt100 thermistor. Current-voltage curves can also be obtained if the voltage-divider circuit is driven by the analog output of the Arduino board: varying the output voltage thus varies the current flowing through the sample. A simple low-pass RC filter should be added, since the analog output is actually 0-5 V pulse-width modulated at 980 Hz and needs to be averaged to produce a real DC voltage.[1,24] The value of the current is measured by the voltage drop across the reference resistor, and the value of the voltage is read directly. The students developed and carefully tested the measurement circuits. They also worked on the sample-thermometer thermalization and built a setup with a large thermal inertia to control the rate of temperature variations: the sample was embedded in a beaker full of glass beads and liquid nitrogen was used to provide cooling power. The students used their setup to study different properties of a semiconductor. They could clearly measure the exponential decrease of resistance with temperature and extract the electronic gap of their sample, a NTC thermistor. They found the reasonable value of 0.22 +/- 0.01 eV. After verifying Ohm's law on a resistor, they performed I-V curves on LED's p-n junction and showed that the value of the threshold voltage presents a temperature dependence of - 2 mV/K consistent with the literature,[25] as shown in Figure 1.

| project | Performed measurements | Sensors used for the project |
|---|---|---|
| Thermoelectric properties | Seebeck coefficient versus temperature for different metals | Arduino analog inputs, voltage amplifier, Pt100 |
| Superconductivity | Resistivity versus temperature | Arduino analog inputs, voltage amplifier, Pt100 |
| Semiconductor | Resistivity versus temperature and hall effect | Arduino analog inputs and outputs, voltage amplifier, Pt100 |
| Ferromagnetism | Magnetization versus field for different ferromagnetic materials | Hall sensors |
| Induction | Induced voltage in a coil for different geometries | Arduino analog input |
| Mechanical properties | Young modulus for different metals | Hall sensor |
| Acoustic | Sound velocity | Electret microphone |
| Acoustic | Sound absorption for different materials | Electret microphone |
| Percolation | Electrical conductivity of a mixture of metal and glass beads | Arduino analog input |
| Scales | Fabrication, calibration and test of reproducibility of a weighing scale | Arduino analog inputs, force sensing resistors |
| Peltier cell | determination of a Peltier cell's parameters | Arduino analog inputs thermocouples and voltage amplifier |

Table 2: list of students' projects.

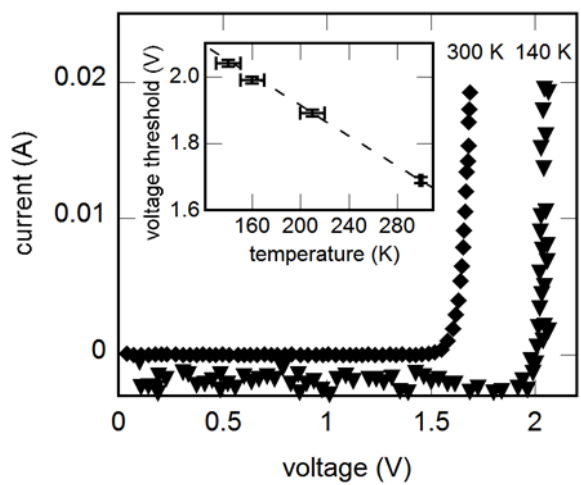

Figure 1: (adapted from a student report) current-voltage dependence of a LED at two different temperatures. The noise in the low temperature data is probably due to a degradation of the LED after being cooled down rapidly with liquid nitrogen. Insert: temperature dependence of the threshold voltage.

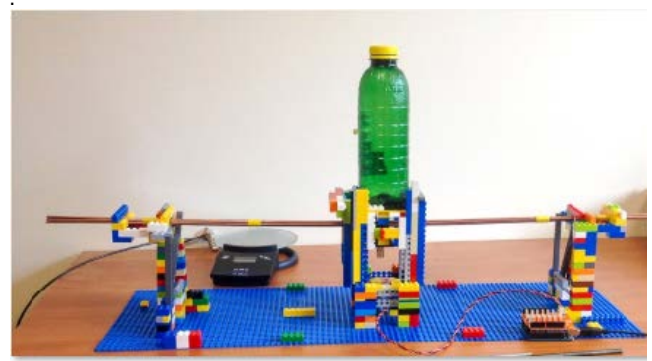

Figure 2: experimental setup of the project studying the deformation of a metallic rod with fixed extremities. The water bottle stands on a platform that is free in the vertical direction. By adding water in the bottle, the weight applied on the rod increases. A magnet is attached at the middle of the rod, and a Hall sensor is fixed on the table below to measure the vertical deformation when the force is increased.

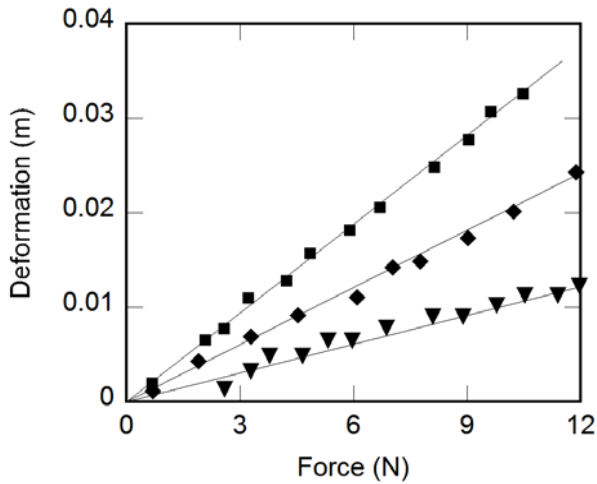

Figure 3: (adapted from a student report) deformation of a copper rod vs applied force at the middle of the rod. Different rod lengths are represented. The slope of the linear fits is related to the Young modulus.

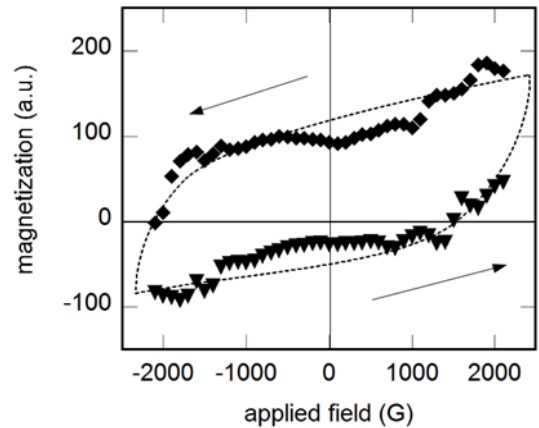

Figure 4: (adapted from a student report) magnetization of a ferrite magnet vs magnetic field. The y-axis actually represents the magnetic field difference (in gauss) measured by the Hall sensor between a run with the sample and a run without the sample. The dashed line is a guide to the eyes. The asymmetry of the curve shows that the applied magnetic field is not large enough to completely reverse the magnetic moment of the sample.

### B. MECHANICAL PROPERTIES OF MATTER

Various sensors can be used to measure a force and a displacement (see Table 1). In the present project, the students chose to measure the deformation of a rod when a load is applied at its middle,[26] as shown in Figure 2. To measure the displacement of the rod, a magnet was fixed at its center, and a hall sensor was fixed on the table below to measure the magnetic field created by the magnet. This setup has a resolution of about 1 mm on a 10 cm range and excellent reproducibility. The value of the measured magnetic field was calibrated as a function of the distance to the magnet. The force was applied using a plastic bottle as a variable weight by varying the quantity of water inside (see the setup on Figure 2). The students developed and optimized their setup and could study the effect of the length and thickness of the rod on the deformation and determine the Young modulus of the metal, see Figure 3. They compared their results for copper and steal with the literature values, and found that their Young modulus value for copper was almost twice the reported value (210 +/- 30 GPa instead of 129 GPa[27]). The students decided to check the nature of the rod metal using the X-ray facility they had used in a previous student lab: the metal was really copper, as advertised. Following a suggestion of their instructor, they annealed their rod and realized that the mechanical properties of the metal were modified.

### C. MAGNETIC PROPERTIES OF MATTER

This projects aims at measuring magnetization versus applied field in various solid materials. A DC magnetic field is applied by approaching a 3 cm-diameter NdFeB magnet. Varying the distance between the magnet and the sample scans the applied magnetic field up to 2000 G, and reversing the magnet gives a -2000 G to +2000 G range. A Hall sensor is positioned close to the sample and measures the total magnetic field, which is the sum of the applied field and the field created by the sample's magnetization. The latter is directly proportional to the magnetization of the sample. The applied field has thus to be removed in order to isolate the magnetization signal. This can be done by comparing an empty run (no sample) to a full run (with a sample), ensuring the reproducibility between runs by using a second Hall sensor to accurately measure the distance to the magnet. The students built their experimental setup in incremental steps, testing various designs. The reproducibility of the magnet position measurement remained the main source of error. In the end, their setup could measure the magnetization of a sample provided the magnetic field produced by it was at least a few gauss at a distance of a few millimeters. This allows to measure ferromagnetic and superconducting materials, but is not sensitive enough to measure weaker paramagnets or diamagnets. The students measured the magnetization of soft iron, the magnetization of a NeFeB magnet and how it is affected after annealing the magnet above 500°C, and they determined the magnetic hysteresis loop of a ferrite magnet (see Figure 4).

## V. THE IMPACT OF THE OPEN LABS ON STUDENTS AND TEACHERS

As we wanted to have a first feedback from the students to gauge their reception of these projects, we sent them a survey consisting of a series of open-ended questions. The goal was not to assess students' knowledge improvement, but to see how their perceptions compare with our objectives. Before the project week, at the beginning of the year, the open projects were proposed as an alternative choice to the more standard focused labs. Among 103 students, 50 chose to follow the open projects, but a limit was set to 24 because of practical constraints. Both open and focused projects were held in parallel with a similar schedule (focused labs have a two-day training period on a National Instrument data acquisition card and a five-day project period), with the same instructors, and for similar students, allowing us to compare the perception and opinions of students following the open projects to those following focused labs. Not all students answered the survey, so we compared 17 students for the open projects and 21 for the focused ones. Table 3 presents the questions and the answers classified in categories. These categories were determined by an a posteriori analysis car-ried out by two authors independently to ensure objectivity. Due to the relatively small number of participants, we will only use these survey results to discuss general trends among students' answers.

| | | open-project students | focused-lab students |
|---|---|---|---|
| **"What is the contribution of this lab week to your scientific formation?"** | | | |
| experimental capacities | "the most important thing I learned is how to interpret data", "I learned to perform experimental work" | 65% | 24% |
| academic knowledge | "knowledge on transition phases", "a better understanding of superconductivity" | 30% | 57% |
| difficulty of doing an experiment | "doing a proper measurement is not easy" | 29% | 14% |
| **"What was the impact of this lab week on you, on a personal basis?"** | | | |
| autonomy, patience and perseverance | "I grew in autonomy", "patience, and not giving up" | 18% | 57% |
| team work | "team work [...] communication, sharing of ideas and point of view" | 53% | 24% |
| work organization | "organization and a better capacity for team work" | 24% | 0% |
| **"What did you particularly appreciate in this lab week?"** | | | |
| autonomy and liberty | the total liberty we had: we had our own room, our own setup, our own topic", "the liberty, being able to try things and fail" | 59% | 38% |
| interaction with teachers | "I appreciated the teachers' trust", "the exchanges with the teachers" | 41% | 19% |
| team work | "group work" | 24% | 19% |
| having five full days | "having a lengthy student lab gives a different point of view" | 0% | 24% |
| **"According to you, what was lacking in this lab week?"** | | | |
| time | "Time?", "more time", "Time. It is a pity that it is so short. I wouldn't have mind having a week more" | 59% | 38% |
| better equipment | "more precise equipment", "better equipment" | 35% | 24% |
| **"Would you like to have a similar lab week again?"** | | | |
| yes | | 82% | 76% |
| no | | 6% | 14% |

Table 3: results of the students' survey (17 answers for the open projects and 21 for the focused projects). Only the most significant answers are reported. The percentages correspond to the fraction of answers that belong to a given category.

To the question "What is the contribution of this lab week to your scientific formation?" the open-project student answers put more emphasis on the experimental process (65%), whereas a majority of focused-lab student answers enters into the theoretical concepts category (57%). This significant difference mirrors the emphasis put on experimental methods rather than theoretical concepts by the open labs.

To the question "What was the impact of this lab week on you, on a personal basis?", team work and work organization is much more often quoted by open-project students than by focused labs ones (53% vs 24% and 24% vs 0%). More surprising is the low mention of autonomy by the open-project students (18%), especially compared to the focused-lab students (57%). However, to the question "What did you particularly appreciate in this lab week?", more than half (59%) of the open-project students enjoyed being autonomous (compared to 39% for the focused-lab students). It is interesting to note that the students appreciated the autonomy during these projects, but did not acknowledge it as an important factor in the previous question.

The interactions with the teachers were also often mentioned: in 41% of their answers, compared to 19% for the focused lab students, whereas the teachers spent as much time with both type of students. In a PBL the role of a teacher is more about mentoring, helping students to reach their own decisions, than teaching in the traditional sense. In the open projects every setup is as new to the teachers as to the students, and it seems that the students were sensible to the different dynamic it creates.

To the question "According to you, what was lacking in this lab week?", time was cited by 59% of the open-project students (compared to 38% for the focused-lab students). The need for better equipment was mentioned in a similar fashion by the open-project (35%) and focused-lab students (24%). This latter result is surprising since the equipment used in the focused labs is often of laboratory quality, and the budget for these labs is in the 10000-euro per-setup ballpark.

Finally and most important, to the question "Would you like to have a similar lab week again?" only a marginal number of students answered "no". No difference could be observed between open-project and focused-lab students.

The five instructors who mentored these labs (among which two authors of the article) were also asked about their teaching experience. All teachers consider that the main upside of these projects is that they constitute a good introduction to the experiment, that it gives students "a better understanding of what measuring means". Among other upsides, autonomy and the fact that students can choose and accomplish their own projects are often quoted.

As downsides, it was noted that the students often focus on the experiment itself to the detriment of the physics at play and the analysis of the data, which could have been lead further in many projects ("students did not push enough their experiment"). Some students seemed to consider that obtaining data was enough to complete a study, and failed to analyze their results as thoroughly as possible to obtain more physical information.

To avoid students frustration in front of poor data, defining clearly the goals of a project appears essential, as is well known for PBL.[9,14] Also, the difficulty of mentoring was noted: "difficult to know how much to help the students, and not doing the project in their stead". The importance of the instructor's role in a PBL activity is well documented.[9,12,13] In our case, we think that the condensed period of time dedicated to the projects increases the importance of this point.

Last but not least, all teachers enjoyed mentoring the open projects, and are willing to do it again the coming year. They appreciated the variety of the projects, the pleasure of the challenge for the teacher, and of discovering new thing. PBL is known to engage students and to increase their motivation; it can also be engaging for the teachers.

## VI. CONCLUSION

New low-cost technology, such as the open-source Arduino microcontrollers and associated sensors, opens the route to simple implementation of project-based physics student labs. This article describes a practical framework for such labs. We demonstrate that within this framework, students can perform pertinent studies of physical phenomena at the level of third-year university curricula even with this low-cost equipment. Our survey on students' and teachers' perceptions suggests that students felt engaged by their projects, discovered experimental physics, and appreciated this intense lab week. The majority of the students mentioned better experimental methods as contribution to their scientific formation. Even though a quantitative assessment of the students' knowledge and skills improvements during these labs was not done, our study suggests that the knowledge gained by the students is less conceptual than in a traditional student lab and more centered on soft skills, such as autonomy and team work.

Beyond this particular PBL framework, the possibility to do physics experiments with a low-cost hardware opens some interesting possibilities. It could be used to develop university physics curricula in emerging countries where limited funding is available to build new labs. It could also be used to let the students perform experiments outside of the university walls. One could imagine a physics curriculum that includes homework with "do-it-yourself" experiments using Arduino boards and simple electronic sensors that could be lent to students, with tasks or challenges corresponding to the lesson of the day. It could also be particularly useful in the case of an online education curriculum and MOOC type of approach. Finally, the development of similar Arduino teachings in other universities could encourage new types of exchanges among physics teachers and students all around the world.

**ACKNOWLEDGEMENTS**: We thank the students who participated to these projects, and Patrick Puzo for welcoming this new teaching in the Magistère de Physique Fondamentale (Université Paris-Sud). We also would like to thank Fabrice Bert, Claire Marrache and Miguel Monteverde for participating as teachers/mentors, and Bertrand Pilette and Christophe Lafarge for their help. This work benefited from the support of the Chair "Physics Reimagined" led by Paris-Sud University and sponsored by AIR LIQUIDE, and also from a grant "pédagogie innovante" from IDEX Paris-Saclay.

\* frederic.bouquet@u-psud.fr
Web site: http://www.PhysicsReimagined.com